\documentclass{easychair}

\usepackage{doc}
\usepackage{todonotes}
\usepackage{subcaption}
\usepackage{enumitem}
\usepackage{float}
\usepackage{lmodern}
\usepackage[T1]{fontenc}

\title{Enhancing Orthopedic Surgical Training With Interactive Photorealistic 3D Visualization}%

\author{
Roni Lekar\inst{1}\thanks{Corresponding author; software, conducting user study, writing - original draft}
\and
Tatiana Gerth\inst{2}\thanks{Software, conducting user study, writing - original draft}
\and
Sergey Prokudin\inst{3, 4}\thanks{Shaping user study, recordings and 3D reconstructions, writing - reviewing}
\and
Matthias Seibold\inst{4}\thanks{Shaping user study, writing - reviewing}
\and
Reto Bürgin
\inst{2}\thanks{Data analysis}
\and
Benjamin Vella\inst{5}\thanks{Organizing participants}
\and
Armando Hoch\inst{5}\thanks{Setting and verifying medical content}
\and
Siyu Tang\inst{3}\thanks{Infrastructure, methodology and supervision}
\and
Philipp Fürnstahl\inst{4}\thanks{Infrastructure, methodology and supervision}
\and
Helmut Grabner\inst{1}\inst{2}\thanks{Feedback, support, methodology and supervision}
}

\institute{
Institute of Computer Science, School of Engineering, Zurich University of Applied Sciences, Winterthur, Switzerland
\email{\{roni.lekar, helmut.grabner\}@zhaw.ch}
\and   
Institute of Data Analysis and Process Design, School of Engineering, Zurich University of Applied Sciences, Winterthur, Switzerland
\email{\{tatiana.gerth, reto.buergin\}@zhaw.ch}
\and  
Vision and Learning Group, ETH Zurich, Zurich, Switzerland
\email{\{sergey.prokudin,siyu.tang\}@inf.ethz.ch}
\and
Research in Orthopedic Computer Science, Balgrist University Hospital, University of Zurich, Zurich, Switzerland
\email{\{matthias.seibold, philipp.fuernstahl\}@balgrist.ch}
\and
Department of Orthopaedics, Balgrist University Hospital, University of Zurich, Zurich, Switzerland
\email{armando.hoch@balgrist.ch}
\and
Faculty of Medicine, University of Zurich, Zurich, Switzerland
\email{benjamin.vella@uzh.ch}
}

\authorrunning{Lekar et al.}

\titlerunning{Enhancing Orthopedic Surgical Training With Interactive Photorealistic 3D Visualization}

\begin{document}

\maketitle

\begin{center}
\url{https://panteroni.github.io/3dsurgerytraining}

\end{center}

\begin{abstract}
Surgical training integrates several years of didactic learning, simulation, mentorship, and hands-on experience. Challenges include stress, technical demands, and new technologies. Orthopedic education often uses static materials like books, images, and videos, lacking interactivity. This study compares a new interactive photorealistic 3D visualization to 2D videos for learning total hip arthroplasty. In a randomized controlled trial, participants (students and residents) were evaluated on spatial awareness, tool placement, and task times in a simulation. Results show that interactive photorealistic 3D visualization significantly improved scores, with residents and those with prior 3D experience performing better. These results emphasize the potential of the interactive photorealistic 3D visualization to enhance orthopedic training.
\end{abstract}


\setcounter{tocdepth}{2}

\newpage

\section{Introduction}
\label{sect:introduction}
Spatial awareness in orthopedic surgery remains challenging, requiring innovative teaching strategies \cite{3dVisualizationMedicalEducation}. Traditional tools like textbooks and videos \cite{Orthobullets, Video-Based-Education} lack interactivity, which is more effective for learning \cite{talebi2022measuring}. While videos outperform textbooks \cite{videosVsBooks}, their fixed angles and visual obstructions highlight the need for interactive modalities.
\newline\newline
Extended reality technologies, including virtual and augmented reality (AR), are emerging in surgical education \cite{visualizationAR, ultrasoundAR, luohong}, particularly in total hip arthroplasty (THA), enhancing accuracy and skill assessment \cite{VR-Effectiveness, VR-Objective-Measurement-of-Skills}. However, AR hardware requirements can limit accessibility. 
\newline\newline
Advances in 3D reconstruction \cite{xie2022neural} now enable photorealistic, interactive models from 2D images via methods like Neural Radiance Fields \cite{nerf} and 3D Gaussian Splatting (3DGS)\cite{gaussianSplatting}, potentially improving procedural understanding in orthopedic training.
\newline\newline
We introduce a novel approach combining photorealistic 3D reconstructions, using 3DGS, with interactive elements to immerse learners in THA surgery. A user study compared this method with traditional video recordings, detailing study design, participant recruitment, and evaluation methods.


\section{Methods}
\begin{figure}[H]
    \centering
\includegraphics[width=1\linewidth]{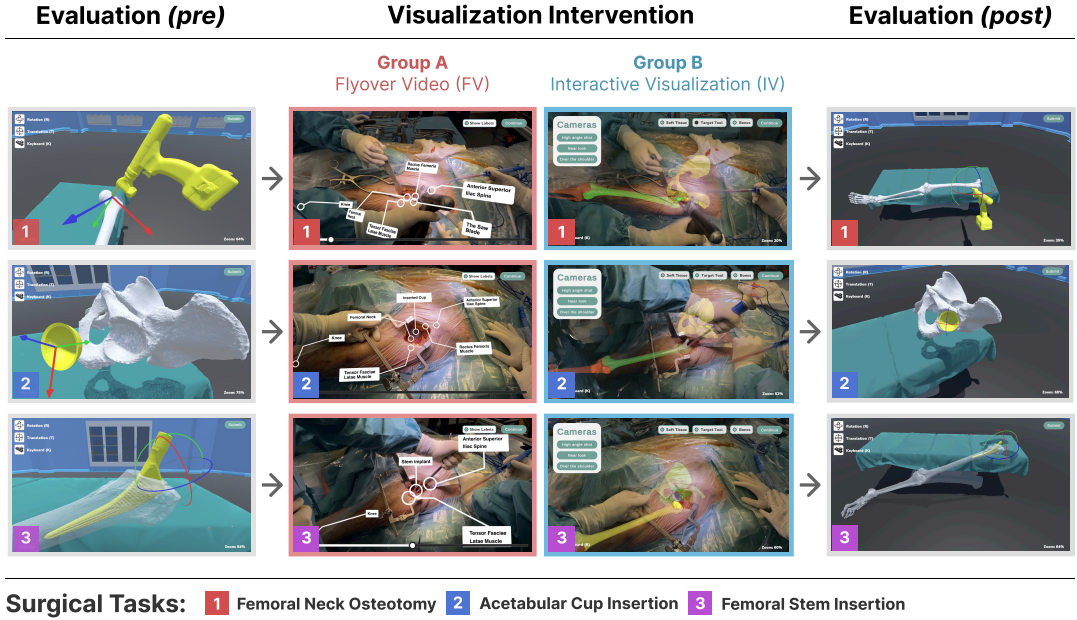}
    \caption{
    The evaluation scene is shown before and after the visualization intervention to measure learning outcomes for each modality. The three surgical tasks evaluated are: Femoral Neck Osteotomy (task 1), Acetabular Cup Insertion (task 2), and Femoral Stem Insertion (task 3). Each participant completed these tasks in sequence. 
IV enhances spatial awareness with \textbf{immersive navigation} for spatial exploration \cite{Spatial-Cognition}, \textbf{gizmo-based transformations} for intuitive tool use \cite{Gizmos}, \textbf{shadows} for depth perception \cite{Shadows}, and \textbf{transparent bones} for clearer tool placement. It also provides unrestricted camera control and layer toggling, addressing the fixed angles of video-based methods.
    }
    \label{fig:flow}
\end{figure}

The study involved 56 participants, including 40 students and 16 residents, with varying THA experience. A randomized controlled trial with a \textit{pre}–\textit{post} design was conducted on a laptop. Participants watched a THA overview video and completed a tutorial to navigate the virtual environment. Knowledge gain was assessed through three THA tasks identified by senior surgeons as indicators of procedural understanding (Figure \ref{fig:flow}).
\newline\newline
Participants were randomly assigned to one of the following two groups:

\begin{description} \item[Group A – Flyover Video (FV):] A non-interactive dynamic video of a specific static scene from the surgical step, with labeled anatomical structures overlaid. \item[Group B – Interactive Visualization (IV):] A photorealistic 3D model \cite{gaussianSplatting} reconstructed from FV footage, enabling free navigation and toggling of anatomical structures.\end{description}

\noindent{A scoring system was developed with a senior surgeon to compare task accuracy, combining position and rotation scores into a final task score, as shown in task 1’s formula:
}
\[
\text{PositionScore} =
\begin{cases}
100 & \text{if } \text{\textit{distPos}} \leq 0.05\text{m} \\
100 \cdot \left(1 - \frac{\text{\textit{distPos}} - 0.05}{1 - 0.05}\right) & \text{if } 0.05\text{m} < \text{\textit{distPos}} < 1\text{m} \\
0 & \text{otherwise}
\end{cases}
\]

\[
\text{RotationScore} =
\begin{cases}
100 & \text{if } \text{\textit{distRot}} \leq 0.2 \\
100 \cdot \left(1 - \frac{\text{\textit{distRot}} - 0.2}{1 - 0.2}\right) & \text{if } 0.2 < \text{\textit{distRot}} < 1 \\
0 & \text{otherwise}
\end{cases}
\]

\[
\text{FinalScore} = \frac{\text{PositionScore} + \text{RotationScore}}{2}
\]
\textit{distPos} is the shortest distance from the saw tip to acceptable bone entry points in meters and \textit{distRot} is the magnitude of the cross product between the normal vectors of the submitted and target saw planes. For the other tasks, rotation scores use rotation axes, but overall it remains consistent with thresholds and linear interpolation.
\noindent{The local ethics committee approved the study. Participants provided informed consent with safeguards for anonymity.}

\newpage
\section{Results and Discussion}
\label{sect:results}

The statistical analysis employed linear mixed models \cite{LinearMixedModel} implemented in R with the lme4 package \cite{lme4}, enabling the examination of:
\begin{enumerate}[label=\roman*, itemsep=0pt]
    \item Whether improvement from \textit{pre} to \textit{post} is greater for IV than FV.
    \item The \textit{post}-group difference, considering possible covariate effects and intra-individual correlations. 
\end{enumerate}
The models used computed scores as the response variable, with task, option (FV vs. IV), stage (\textit{pre} vs. \textit{post}), resident surgeon, and 3D experience as explanatory variables. Random intercepts accounted for intra-individual correlations. Bootstrap methods 
computed $p$-values. 
Regression coefficients ($\beta$) represent the estimated score difference when an explanatory variable changes by one unit,
with $p$-values below 0.05 indicating significance at the 5\% level.\newline\newline
Comparing \textit{post} scores, the IV group performed slightly better, but the difference was not significant ($\beta = 4.230, p = 0.062$). 
For \textit{pre} scores, they tended to score lower than the FV group across all tasks ($\beta = 4.200, p = 0.062$), despite random group assignment. This imbalance likely reflects random variation, not systematic bias. 
Based on these results, our analysis focuses on the \textit{pre}-to-\textit{post} score changes - \textbf{the learning effect}; reflecting improvements in performance after an intervention - rather than just the \textit{post} results.\newline\newline
Across all tasks, IV resulted in a significantly greater learning effect than FV ($\beta = 8.420, p = 0.007$). Residents performed better overall ($\beta = 4.030, p=0.012$), likely due to their subject familiarity, and those with 3D knowledge outperformed others ($\beta = 5.490, p=0.007$), consistent with previous studies \cite{Arthroscopy-Games-Correlation}. Regular use is crucial for tool effectiveness. Data showed minimal impact of time on task completion, with difficulty and familiarity playing larger roles. 
\begin{figure}[H]
    \centering
 \includegraphics[width=1\linewidth]{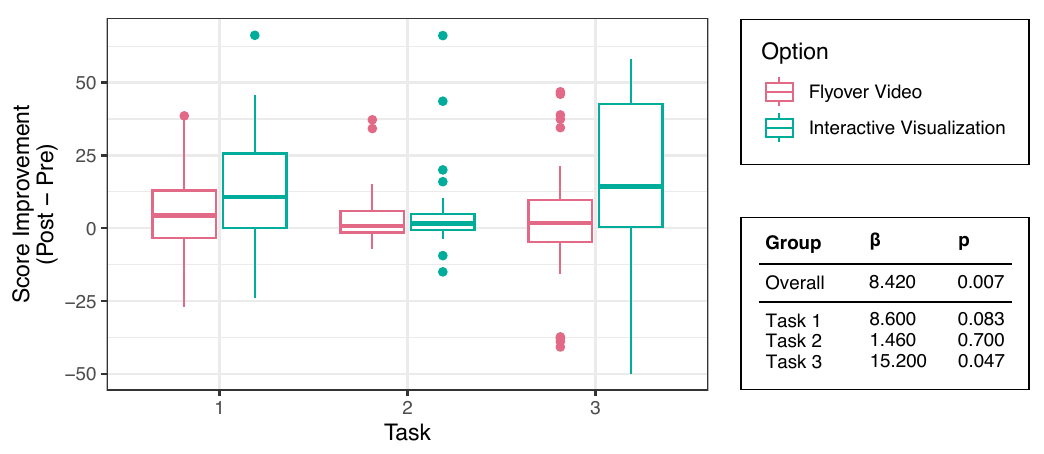}
    \caption{
    Comparison of the learning effect within tasks between FV and IV, with the boxplot illustrating the distribution and mean of the learning effect. The table on the right illustrates how IV outperformed FV. The $\beta$ and $p$ values refer to the estimated extent to which the learning effect was higher for IV and the corresponding $p$-value, derived from a linear mixed model. Significant learning effects are observed overall ($p = 0.007$) and for Task 3 ($p = 0.047$), highlighting the advantage of IV.
    }
    \label{fig:results}
\end{figure}
\noindent{The comparison of learning effects for each task, as shown in Figure \ref{fig:results}, highlights the overall advantage of IV over FV. Task 1 showed a trend toward greater learning effect for IV. The smaller visible anatomy section makes interpreting the osteotomy's impact challenging, but IV simplifies this task significantly. Task 2 exhibited minimal differences between groups, reflecting the task's simplicity (high \textit{pre} and \textit{post} scores) and clear geometric structure. Notably, task 3 revealed a significant advantage for IV, emphasizing the importance of full femur visualization for accurate judgment of 3D position.}
\newline\newline
\noindent{Although task 3 was the only one with a statistically significant result at the 5\% level, this does not contradict the overall results, as task-specific tests use smaller sample sizes. These findings reveal that IV provides a significantly greater learning effect than existing tools like FV \cite{Orthobullets}.}

\section{Conclusion and Future Work}
\label{sect:conclusion-future-work}

The results showed that 
overall the new IV method led to significantly bigger improvement than the FV method ($\beta = 8.420, p = 0.007$). They also indicated that residents and participants with 3D experience tended to achieve better outcomes. These preliminary results show potential in the new method for better skill retention than traditional approaches, but further investigation is needed. The limited sample size may affect the robustness of these findings.
Our limitations include the high cost of recording surgeries, requiring rental of specimens and operating rooms, while adhering to strict hygiene standards. Additionally, the objects must remain still to avoid blurriness in the model. Nevertheless, the study’s findings could enable a library of surgeries featuring IVs of key steps, accessible across various platforms.

\section{Acknowledgments}
\label{sect:acks}

We thank the medical students, residents and orthopedic surgeons who participated. This study was conducted within the national ”Proficiency” research project funded by Innosuisse.


\label{sect:bib}
\bibliographystyle{plain}
\bibliography{easychair}


\newpage
\appendix
\section*{Appendix}
\addcontentsline{toc}{section}{Appendix}

\section{Additional Task Formulas}

In the main text, we presented the formula used to compute performance in task 1. For completeness, we include below the formulas used for tasks 2 and 3.\\
The scoring formula for task 2 (cup insertion) is:

\[
\text{CupPosition} =
\begin{cases}
100 & \text{if } \text{distPos} \leq 0.09m \\
100 \times \left(1 - \frac{\text{distPos} - 0.09}{1 - 0.09}\right) & \text{if } 0.09m < \text{distPos} < 1m \\
0 & \text{otherwise}
\end{cases}
\]

\[
\text{CupRotation1} =
\begin{cases}
100 & \text{if } \text{distRot1} \leq 8^\circ \\
100 \times \left(1 - \frac{\text{distRot1} - 8}{45 - 8}\right) & \text{if } 8^\circ < \text{distRot1} < 45^\circ \\
0 & \text{otherwise}
\end{cases}
\]
\[
\text{CupRototaion2} =
\begin{cases}
100 & \text{if } \text{distRot2} \leq 11^\circ \\
100 \times \left(1 - \frac{\text{distRot2} - 11}{45 - 11}\right) & \text{if } 11^\circ < \text{distRot2} < 45^\circ \\
0 & \text{otherwise}
\end{cases}
\]

\[
\text{FinalScore} = \frac{\text{CupPosition} + \text{CupRotation1} + \text{CupRotation2}}{3}
\]
\textit{distPos} is the Euclidean distance (in meters) between the submitted cup and the target cup.\newline
\textit{distRot1} is the difference (in degrees) between the submitted and target red/x-axis angles.\newline
\textit{distRot2} is the difference (in degrees) between the submitted and target blue/z-axis angles.
\newline\rule{\linewidth}{0.4pt}
The score formula for task 3 (stem insertion):

\[
\text{StemPosition} =
\begin{cases}
100 & \text{if } \text{distPos} \leq 0.06m \\
100 \times \left(1 - \frac{\text{distPos} - 0.06}{1 - 0.06}\right) & \text{if } 0.06m < \text{distPos} < 1m \\
0 & \text{otherwise}
\end{cases}
\]

\[
\text{StemRotation} =
\begin{cases}
100 & \text{if } \text{distRot} \leq 4^\circ \\
100 \times \left(1 - \frac{\text{distRot} - 4}{45 - 4}\right) & \text{if } 4^\circ < \text{distRot} < 45^\circ \\
0 & \text{otherwise}
\end{cases}
\]

\[
\text{FinalScore} = \frac{\text{StemPosition} + \text{StemRotation}}{2}
\]
\textit{distPos} is the absolute difference (in meters) between the distance from the trochanter minor to the submitted stem end and the distance from the trochanter minor to the target stem end.\newline
\textit{distRot} is the angular difference (in degrees) between the submitted and target red/x-axis angles.

\section{Questionnaire and Results}

Due to space constraints, we could not include details of the post-task questionnaire in the main paper. This section provides an overview of the study procedure and summarizes both quantitative and qualitative results gathered from participant feedback.

\subsection{Study Procedure and Questionnaire Design}

After completing the task in their assigned condition (FV or IV), participants filled out a post-task questionnaire. This included the NASA Task Load Index (NASA-TLX) to assess perceived workload, the System Usability Scale (SUS) to evaluate usability, and several custom questions, including open-ended prompts for qualitative feedback.\\
\\
Following the questionnaire, participants were introduced to the alternative visualization method used by the other group. They were then asked for verbal feedback comparing both approaches.

\subsection{Quantitative Results}

Participants’ perceived workload was assessed using the NASA-TLX questionnaire, with equal weighting across all six dimensions. The average workload score was 7.3 (on a 1–20 scale), indicating a moderate workload. This may reflect the challenge of performing tasks in a 3D environment, as 78.6\% of participants reported limited or no prior experience with 3D systems.\\
\\
Usability was evaluated using the System Usability Scale (SUS). The FV group scored an average of 65.98, while the IV group scored 66.61 (on a 0–100 scale, where higher values indicate better usability). Some UI elements were found to be occasionally confusing, and we are aware of these issues; future iterations will focus on streamlining the interface to enhance clarity and overall usability. \\ 
\\
Across both groups, most participants enjoyed the simulation (64\%) and believed it could enhance their current study materials (75\%). Participants using the IV were also more likely to report that they would use the system regularly (61\% vs. 43\% for the FV group).
\\
\\
To explore how our tool could best be delivered in the future, we asked participants about their preferred type of application to use this tool. This helped us understand which formats are most appealing to different user groups.

\begin{figure}[H]
    \centering
        \centering
        \includegraphics[width=0.9\textwidth]{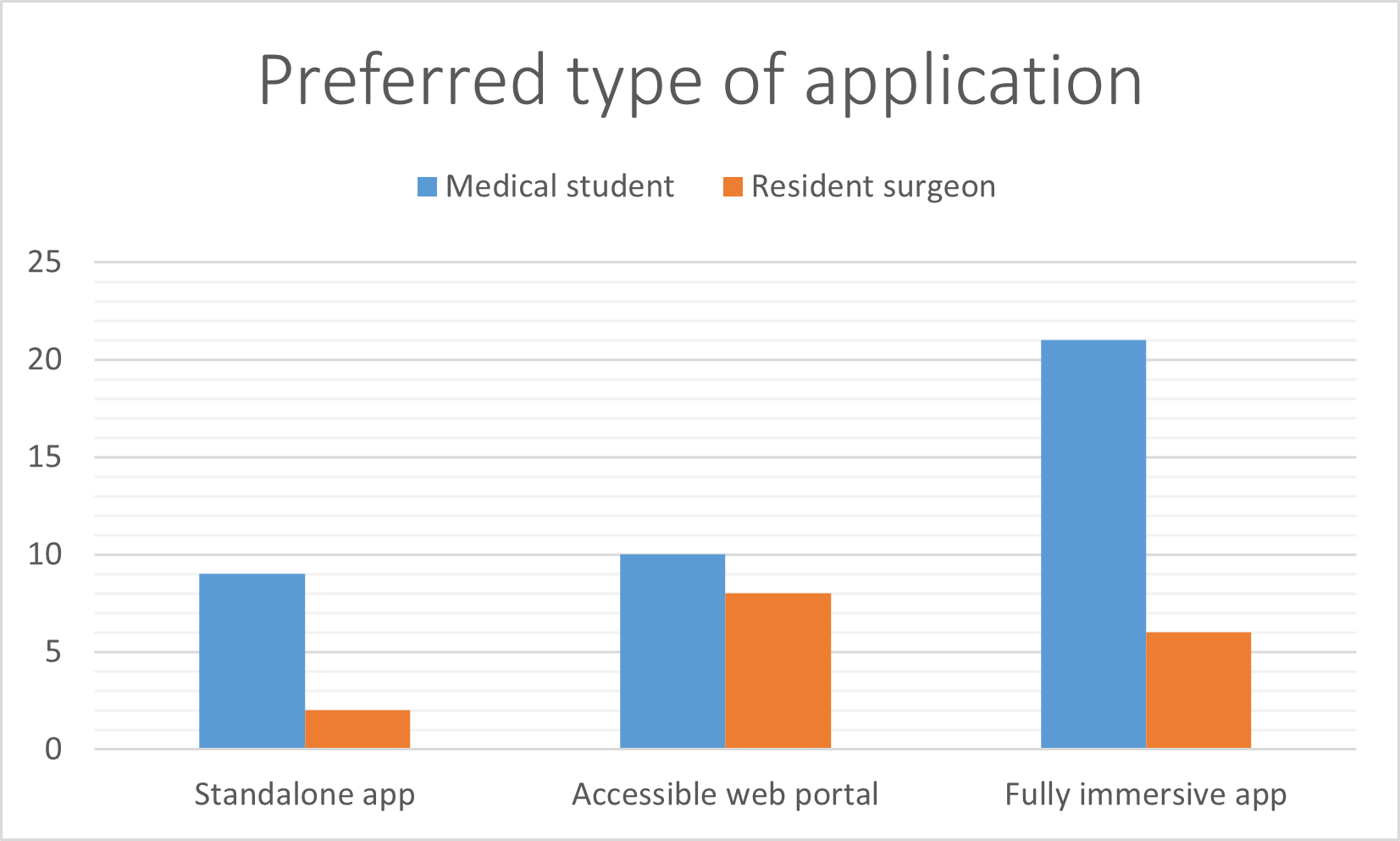}

    \caption{Responses grouped by profession: medical students and resident surgeons. \\\\ The application types were defined as follows: \\\textbf{Standalone app:} A locally installed application requiring no internet connection. \\ \textbf{Accessible web portal:} A browser-based application usable without installation. \\ \textbf{Fully immersive app:} A VR/AR-based application offering an immersive learning experience.}
    \label{fig:preferredApplicationType}
\end{figure}

Overall, the majority favored a fully immersive app, followed by an accessible web portal, while standalone apps were least preferred (Figure~\ref{fig:preferredApplicationType}).

When broken down by profession,
medical students showed a strong preference for immersive apps, while resident surgeons clearly favored web-based formats. This suggests that while students are drawn to more engaging, innovative solutions, more experienced users tend to prioritize practicality and ease of access.

\subsection{Qualitative Results}

All participants expressed a clear preference for the IV over the FV. Interview feedback highlighted that the interactive version felt more engaging and informative, allowing for a more active and self-directed learning experience.\\
\\
In general, participants appreciated the 3D visualization capabilities of the tool, particularly the ability to view angles that would not be accessible in a real surgical setting. Many found it to be a valuable introduction to total hip arthroplasty, especially for first-time learners. 

However, several users found navigation challenging and suggested additional fixed viewpoints, as well as options like touchscreen support or controller-based interaction, to improve orientation. Some participants suggested including animations of specific surgical steps—such as the femoral neck cut—to improve their understanding of the procedure. Additionally, the inclusion of instant feedback during tasks was mentioned as a key improvement to support learning and self-assessment.

\section{Design Decisions and Implementation Details}

\textbf{3D Transformation and Rotation Gizmos.} Similarly to other 3D software, we placed a gizmo around the medical tools. Users can choose between the transformation and rotation (in 3D software they could also choose the scale gizmo, however this isn't relevant in our case, as the tools are fixed with their size). At each step, participants could place a single tool in the correct position.
Red vectors or spheres represent the X-axis, green indicates the Y-axis, and blue corresponds to the Z-axis.
When a user hovers over an axis, it becomes thicker to indicate it is selectable. Upon selection and movement, the axis changes color to yellow to confirm interaction.

\textbf{Casting and Receiving Shadows.} There are shadows of the bones that can be used to identify the depth and 3d position in space. Many participants noted that it helped them understand where in space the stems were compared to the bones. Shadows of the medical tools and the bones are appearing on the floor and on the bed of the operating room as well as on top of each other. For example, the shadow of the saw could be seen on the bone. 

\textbf{Material Transparency}
Some bones, such as the femur in the third step (stem implant) are transparent so the user could see the step through the bone and position it in the best possible way.

\end{document}